\documentclass[aps, prl, superscriptaddress, ctexart, nofootinbib, twocolumn]{revtex4}

\usepackage{graphicx}
\usepackage{latexsym}

\usepackage{color}

\def\be{\begin{equation}}
\def\ee{\end{equation}}
\def\bea{\begin{eqnarray}}
\def\eea{\end{eqnarray}}

\begin{document}

\title{Evidence for bouncing evolution before inflation after BICEP2}

\author{Jun-Qing Xia}
\affiliation{Key Laboratory of Particle Astrophysics, Institute of High Energy Physics, Chinese Academy of Science, P.O.Box 918-4, Beijing 100049, P.R.China}

\author{Yi-Fu Cai}
\affiliation{Department of Physics, McGill University, Montr\'eal, QC, H3A 2T8, Canada}

\author{Hong Li}
\affiliation{Key Laboratory of Particle Astrophysics, Institute of High Energy Physics, Chinese Academy of Science, P.O.Box 918-4, Beijing 100049, P.R.China}
\affiliation{National Astronomical Observatories, Chinese Academy of Sciences, Beijing 100012, P.R.China}

\author{Xinmin Zhang}
\affiliation{Theoretical Physics Division, Institute of High Energy Physics, Chinese Academy of Science, P.O.Box 918-4, Beijing 100049, P.R.China}

%\date{\today}

\begin{abstract}
The BICEP2 collaboration reports a detection of primordial cosmic microwave background (CMB) B-mode with a tensor-scalar ratio $r=0.20^{+0.07}_{-0.05}$ (68\% C.L.). However, this result has a tension with the recent Planck limit, $r<0.11$ (95\% C.L.), on constraining inflation models. In this Letter we consider an inflationary cosmology with a preceding nonsingular bounce which gives rise to observable signatures on primordial perturbations. One interesting phenomenon is that both the primordial scalar and tensor modes can have a step feature on their power spectra, which nicely cancels the tensor excess power on the CMB temperature power spectrum. By performing a global analysis, we obtain the 68\% C.L. constraints on the parameters of the model from the Planck+WP and BICEP2 data together: the jump scale $\log_{10}(k_{\rm B}/{\rm Mpc}^{-1})=-2.4\pm0.2$ and the spectrum amplitude ratio of bounce-to-inflation $r_B\equiv P_{\rm m} / A_{\rm s} = 0.71\pm0.09$. Our result reveals that the bounce inflation scenario can simultaneously explain the Planck and BICEP2 observations better than the standard $\Lambda$CDM model, and can be verified by the future CMB polarization measurements.
\end{abstract}

%\pacs{98.80.Es, 98.80.Cq}

\maketitle

%Introduction==========================================================
%\section{Introduction}\label{Int}

%Inflation, as proposed in early 1980s to understand the initial conditions in the hot big bang cosmology \cite{Guth:1980zm} (see also refs. \cite{Starobinsky:1980te, Fang:1980wi, Sato:1980yn} for early works), has become the most successful paradigm of describing the very early universe. Its prediction of a nearly scale-invariant primordial power spectrum of curvature perturbations has been verified to high precision by the observations on the CMB temperature anisotropies in recent years \cite{Ade:2013zuv}. Inflation also predicted a nearly scale-invariant primordial power spectrum of tensor perturbations, which can seed the B-mode polarization as detected by the BICEP2 experiment.

{\it Introduction.---}Recently, the BICEP2 collaboration announced the detection of primordial B-mode polarization on the CMB. This significant measurement implies that, if all the B-mode polarization signals are contributed by primordial gravitational waves, the corresponding tensor-to-scalar ratio is constrained as \cite{Ade:2014xna}
\begin{equation}
 r = 0.2 ^{+0.07}_{-0.05} ~ (68\%~{\rm C.L.})~.
\end{equation}
This profound discovery has a series of significant implications on very early universe models \cite{Kehagias:2014wza, Lizarraga:2014eaa, Brandenberger:2014faa, Moss:2014cra, Contaldi:2014zua}.
%However, this result is in tension with the standard slow-roll inflation models. And the corresponding excess power in the CMB temperature power spectrum has not been observed by the Planck experiment $r<0.11$ (95\% C.L.) \cite{Ade:2013zuv}.
However, this result has a tension with the recent Planck limit, $r<0.11$ (95\% C.L.) on standard inflation models since the excess power in the CMB temperature power spectrum was not observed by the Planck experiment \cite{Ade:2013zuv}.

In order to lessen the pressure on inflation models and the tension with Planck data, we consider an important extension of inflationary cosmology, which may introduce a nonsingular bounce to connect a contracting phase of the universe with the inflationary stage. It is well known that the big bang singularity issue can be avoided in the framework of bouncing cosmologies \cite{Mukhanov:1991zn, Cai:2007qw}. The scenario of bounce inflation has been applied to suppress CMB anisotropies on large angular scales \cite{Piao:2003zm}. By virtue of the effective field description, it can be achieved by matter fields with the null energy condition violation, such as the quintom bounce \cite{Cai:2008qb, Cai:2008ed}, in which an explicit matter-bounce inflation scenario was obtained with the inflationary epoch being preceded by a contracting phase dominated by the pressureless dust matter. This scenario was also realized in the frame of loop quantum cosmology (namely see ref. \cite{Mielczarek:2010bh} and references therein).

In this Letter, we aim at searching for key observational signals for the bounce inflation scenario which are expected to be sensitive to cosmological CMB measurements. Specifically, we perform an estimate on the power spectrum of primordial gravitational waves generated in the matter-bounce inflation scenario and find that its amplitude undergoes a jump feature at a critical length scale. A similar property was also found in the power spectrum of primordial curvature perturbation as pointed out in ref. \cite{Liu:2010fm}. Using the Planck and BICEP2 data, we perform a global analysis on this bounce inflation scenario and find that it can better interpret the recent CMB observations when compared with the $\Lambda$CDM.

%Method and Current Observations=======================================
%\section{Formalism of a bounce inflation Scenario}

{\it Formalism.---}We begin with a brief discussion of primordial perturbations in the frame of a flat FRW Universe. The relic gravitational waves generated in very early universe is a basic prediction in the modern cosmology \cite{Grishchuk:1974ny, Starobinsky:1979ty}. A standard process of generating primordial power spectrum suggests that, metric fluctuations initially emerge inside a Hubble radius, and then leave it in a primordial epoch, and finally reenter at late times \cite{Mukhanov:1990me}.
The dynamics of primordial gravitational waves is convenient to be investigated by tracking a Fourier mode $v_k$ along the cosmic evolution. In the context of General Relativity, the corresponding equation of motion in the Fourier space is given by
\begin{eqnarray}\label{eom}
 v_k''+(k^2-\frac{a''}{a})v_k=0~,
\end{eqnarray}
where $a$ is the scale factor of the universe and the prime denotes the derivative with respect to the comoving time $\eta\equiv \int dt/a$. Specifically, the scale factor often scales as $a(t)=a_B({t}/{t_B})^{1/\epsilon}$, where the subscript ``$B$" denotes any reference time which will be referred as the bouncing point later. Note that $\epsilon\equiv -\dot{H}/H^2$ is physically associated with the background dynamics, namely, it represents for the slow roll parameter during inflation and equals to $3(1+w)/2$ for other cosmic evolutions with $w$ being the regular equation-of-state parameter. Using comoving time, one derives $a(\eta) =a_B({\eta}/{\eta_B})^{1/(\epsilon-1)}$ for a constant $\epsilon$, and hence, the comoving Hubble rate is given by ${\cal H}\equiv a'/a = 1/(\epsilon-1)\eta$. For instance, for inflation with $\epsilon\ll1$ there is $|{\cal H}| \simeq |1/\eta|$; for a pressureless matter dominated phase with $w=0$ (and thus $\epsilon=3/2$), then $|{\cal H}| \simeq |2/\eta|$. Moreover, there is ${a''}/{a}={(\nu^2-{1}/{4})}/{\eta^2}$, with $\nu=\pm{(\epsilon-3)}/{(2\epsilon-2)}$.
%\end{eqnarray}

We assume cosmological perturbations originate from vacuum fluctuations, which suggests,
%\begin{eqnarray}\label{inicond}
 $v_k^{i}\simeq \exp{(-i\int^\eta kd\tilde\eta)}/\sqrt{2k},$
%\end{eqnarray}
when $|k\eta|\gg1$. This is consistent with the asymptotic solution to Eq. (\ref{eom}) when the last term ${a''}/{a}$ is negligible.
%The mechanism of generating primordial perturbations requires that the absolute value of the comoving time is enough large which can only be achieved in a contracting or an inflationary setup.
Another asymptotic solution to Eq. (\ref{eom}) can be derived in terms of the Bessel function,
%\begin{eqnarray}\label{sollead}
 $v_k \sim \eta^{{1}/{2}} [ c(k)\eta^{-|\nu|} ],$
%\end{eqnarray}
at super-Hubble scales with $|k\eta|\ll1$. Now we match these two asymptotic solutions at the moment of Hubble crossing $|k\eta|\sim1$, and then obtain the tensor mode on super-Hubble scales as
\begin{eqnarray}\label{solution}
 v_k(\eta) \simeq \frac{1}{\sqrt{2k}} (k\eta)^{\frac{1}{2}-|\nu|}~.
\end{eqnarray}
From the definition of the power spectrum $P_T\equiv \frac{4k^3}{\pi^2}|\frac{v_k}{a}|^2$, one easily learns that the scale invariance requires $|\nu|=3/2$ which has to be achieved in a period of matter contraction \cite{Wands:1998yp, Finelli:2001sr} or by inflation. However, the comoving Hubble rate evolves as $|{\cal H}| \simeq |2/\eta|$ during matter contraction while takes another form $|{\cal H}| \simeq |1/\eta|$ during inflation. As a result, if there is a matter contraction before inflation, the amplitude of the power spectrum for primordial gravitational waves would undergo a jump around the scale $k_B$ comparable to the bounce scale. A detailed calculation reveals that $P_T = {H^2}/{2\pi^2}$ when $k<k_B$ while $P_T = {2H^2}/{\pi^2}$ when $k\geq{k}_B$ for the model of matter-bounce inflation.

In analogue with the method developed in \cite{Liu:2010fm}, we phenomenologically parameterize the power spectrum for primordial tensor perturbations as follows,
\begin{eqnarray}\label{parametrize_T}
 P_{T}=P_T^m+\frac{P_T^i-P_T^m}{2}\left\{1+\tanh\left[T_B\log_{10}\left(\frac{k}{k_B}\right)\right]\right\}~,
\end{eqnarray}
with $P_T^m \equiv {H^2}/{2\pi^2}$ and $P_T^i \equiv {2H^2}/{\pi^2}$ being introduced. In addition, the power spectrum of primordial curvature perturbation can be parameterized as
\begin{eqnarray}\label{parametrize_S}
 P_{\zeta}=P_{m}+\frac{P_{\zeta,i}-P_{m}}{2}\left\{1+\tanh\left[T_B\log_{10}\left(\frac{k}{k_B}\right)\right]\right\}~.
\end{eqnarray}
Particularly, $P_{\zeta,i}={H^2}/{8\pi^2\epsilon}$ is the power spectrum during inflation and $P_{m}$ is the spectrum before the bounce which is required to be less than $P_{\zeta,i}$.

\begin{figure*}[t]
\begin{center}
\includegraphics[scale=0.35]{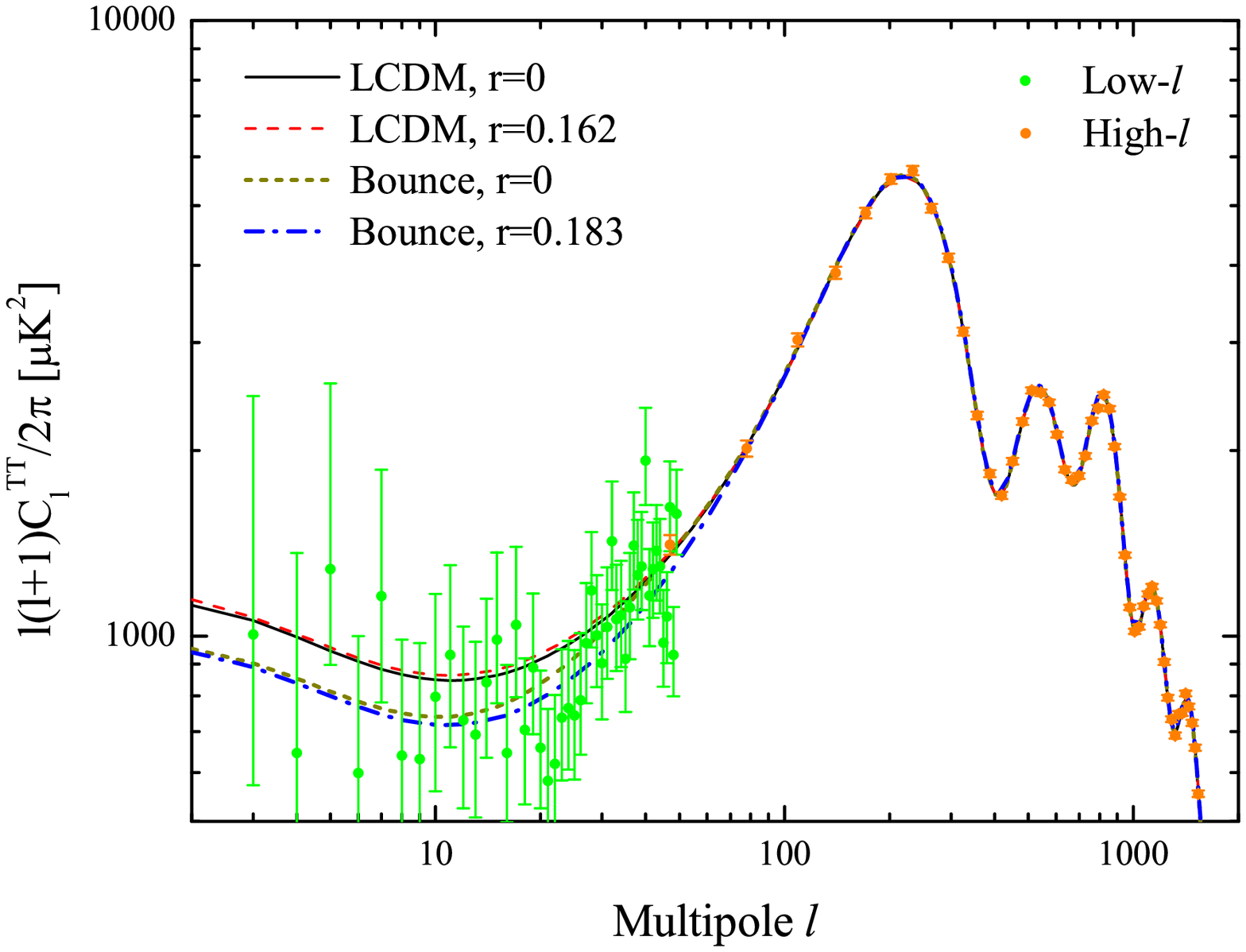}
\includegraphics[scale=0.35]{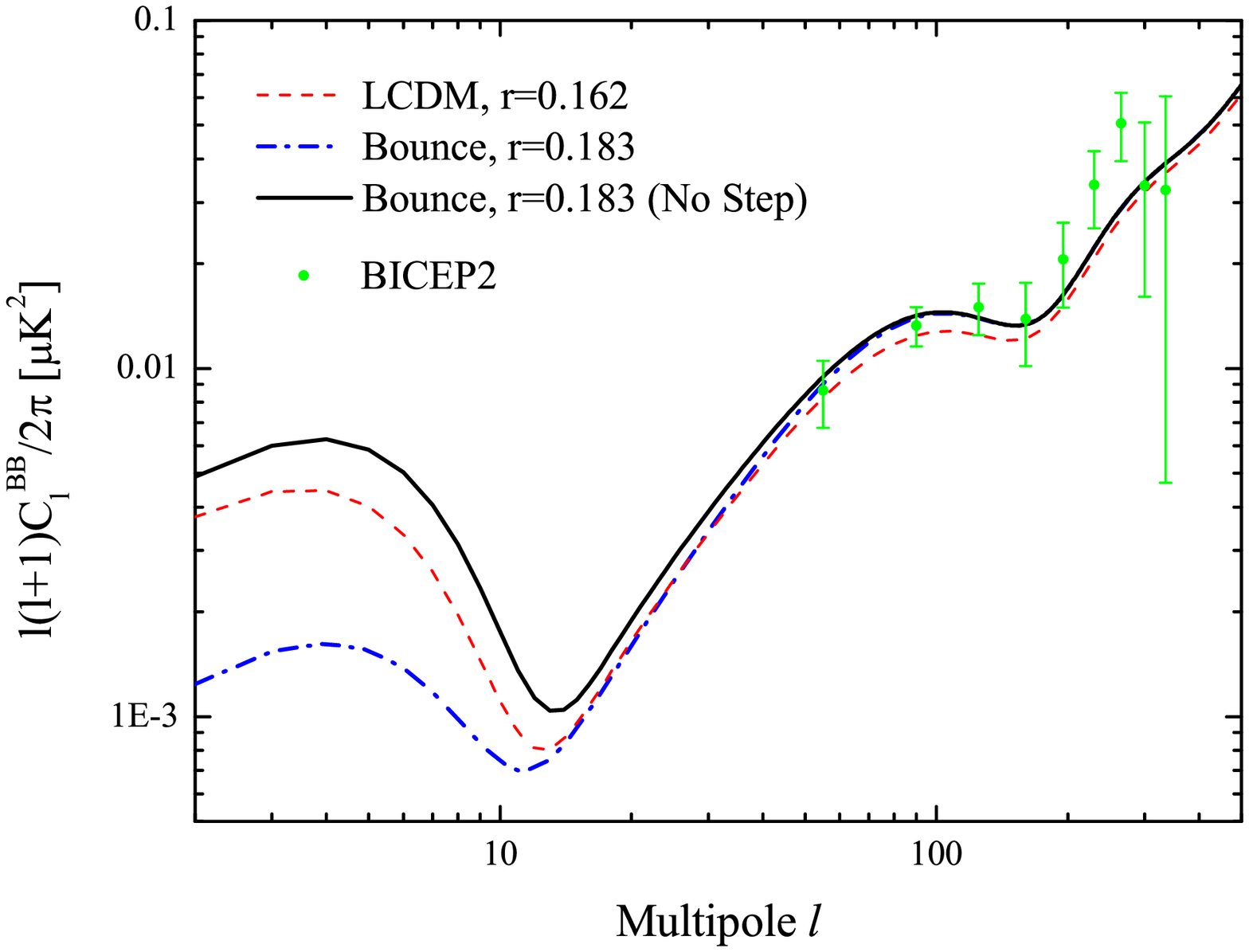}
\caption{Theoretical CMB power spectra for the best fit $\Lambda$CDM models and bounce inflation models, as well as the Planck and BICEP2 observational data. {\it Left}: The CMB temperature power spectra for four best fit models: $\Lambda$CDM models and bounce inflation models with and without using the BICEP2 data. {\it Right}: The CMB BB power spectra for the best fit $\Lambda$CDM and bounce inflation models when using the BICEP2 data. The black solid line denotes the best fit bounce inflation models without the step feature. \label{figure:bestCMB}}
\end{center}
\end{figure*}

As in usual, the power spectrum during inflation can be parameterized as $P_{\zeta,i}=A_s k^{n_s-1}$, in which $A_s$ and $n_s$ are the amplitude and the spectral index correspondingly. Since primordial density fluctuations rely on the model parameters during the bounce, the amplitude of its power spectrum before the bounce can be any arbitrary value lower than that during inflation \cite{Cai:2011zx, Cai:2008qw} \footnote{The equations of motion for primordial density perturbations is similar to Eq. (\ref{eom}) except that the scale factor $a$ is replaced by another background parameter which relies on the specific bounce mechanism, and hence, we treat the spectrum amplitude of density perturbations generated before the bounce to be free.}. Therefore, the observational constraint on primordial density perturbations is pretty loose \cite{Liu:2010fm}. Similar to the analysis of primordial gravitational waves, one can introduce a bounce-to-inflation ratio of power spectrum, $r_B\equiv P_m/A_{s}$ to characterize the spectrum obtained before the bounce. However, for primordial tensor fluctuations, their dynamics only depend on the evolution of the scale factor and hence, once we have determined the background evolution, the power spectrum of primordial gravitational waves can be fixed. Moreover, the parameter $k_B$ denotes the occurrence scale of the jump feature in the power spectrum (in unit of ${\rm Mpc}^{-1}$), and $T_B$ depicts the slope of this jump and thus is associated with the bounce duration. Apparently, these three parameters are highly correlated. We try to constrain them simultaneously, but the results are not good enough, especially when using Planck data alone. Therefore, in our numerical calculations we fix $T_B=5$ which is the best fit value we obtain from Planck+WP+BICEP2 data and constrain the other two parameters.

%Numerical method===============================================================
%\section{Constraints from observational Data}

{\it Results.---}We perform a global fitting using the CosmoMC package \cite{cosmomc}, a Markov Chain Monte Carlo code, which has been modified to calculate the theoretical CMB power spectra in the bounce inflation scenario. We assume adiabatic initial conditions and a flat universe. We vary the following cosmological parameters ($\Omega_bh^2$, $\Omega_ch^2$, $\tau$, $\Theta_s$, $n_s$, $A_s$, $r$), where $\Omega_bh^2$ and $\Omega_ch^2$ are the baryon and cold dark matter densities, $\tau$ is the optical depth to reionization, $\Theta_s$ is the ratio (multiplied by 100) of the sound horizon at decoupling to the angular diameter distance to the last scattering surface, $n_s$ is the spectral index, $r$ is the tensor to scalar ratio of the power spectrum and $A_s$ is the primordial amplitude at the pivot scale $k_0=0.05{\rm Mpc}^{-1}$. Furthermore, we have two more parameters $k_B$ and $r_B$ which are related to the bounce model.

Particularly, we use the low-$\ell$ and high-$\ell$ CMB temperature power spectrum data from the Planck with the low-$\ell$ WMAP9 polarization data (Planck$+$WP). We marginalize over the nuisance parameters that model the unresolved foregrounds with wide priors. For the BICEP2 data, we use their BB power spectrum into our analyses.

\begin{table}[t]
\caption{The $\chi^2$ values for different best fit models from different data combinations. $\chi^2$(P,low-$\ell$) is for the Planck low-$\ell$ TT spectrum only, $\chi^2$(P) is for the Planck+WP data, and $\chi^2$(B) is for the BICEP2 BB spectrum.}\label{chi2values}
\begin{center}
\begin{tabular}{lccc}
\hline \hline

Model & $\chi^2$(P,low-$\ell$) & $\chi^2$(P) & $\chi^2$(B) \\

\hline
$\Lambda$CDM, $r=0$ & $-6.7$ & $9805.7$ & $56.0$ \\
$\Lambda$CDM, $r=0.162$ & $0.7$ & $9814.3$ & $9.0$ \\
Bounce, $r=0$ & $-10.9$ & $9804.3$ & $53.7$ \\
Bounce, $r=0.183$ & $-9.2$ & $9805.6$ & $7.0$ \\

%$\Lambda$CDM, $r=0$ & $-6.7$ & $7797.9$ & $2014.5$ & $92.6$ \\
%$\Lambda$CDM, $r=0.162$ & $0.7$ & $7800.0$ & $2013.6$ & $10.7$ \\
%Bounce, $r=0$ & $-10.9$ & $7800.3$ & $2014.9$ & $91.2$ \\
%Bounce, $r=0.176$ & $-9.2$ & $7800.5$ & $2014.3$ & $7.9$ \\

\hline \hline
\end{tabular}
\end{center}
\end{table}

In table \ref{chi2values} we list the minimal $\chi^2$ values for different cosmological models from different data combinations. In the $\Lambda$CDM, the model with $r=0$ is consistent with the Planck TT power spectrum, $\chi^2(P)=9805.7$, but is strongly ruled out by the BICEP2 data, $\chi^2(B)=56.0$. When including tensor fluctuations in the calculation, the $\chi^2$ value of the best fit model from the BICEP2 data significantly decreases to $\chi^2(B)=9.0$. The BICEP2 data strongly favor a non-zero amplitude of the primordial tensor power spectrum, namely the 68\% C.L. limit is $r=0.162\pm0.034$. This result is consistent with that from the BICEP2 collaboration \cite{Ade:2014xna}. However, the non-zero $r$ model will bring the extra power on CMB low-$\ell$ temperature power spectrum, which leads to the worse fit to the Planck data, especially to the low-$\ell$ data, as shown in the left panel of figure \ref{figure:bestCMB}. Therefore, the standard $\Lambda$CDM model can not simultaneously fit to the Planck and BICEP2 data very well, due to the excess power on CMB TT spectrum at large scales.

\begin{figure}[t]
\begin{center}
\includegraphics[scale=0.4]{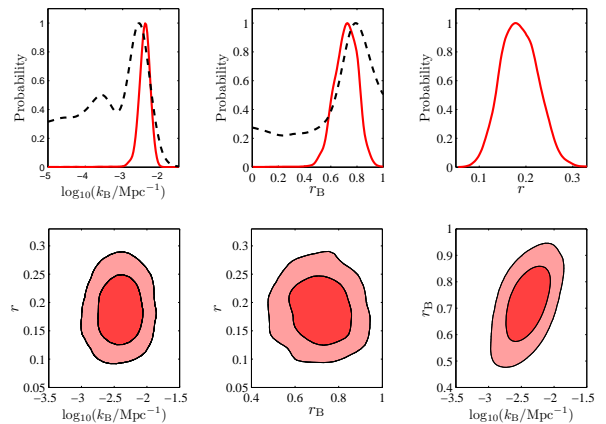}
\caption{One and Two dimensional constraints on the parameters of bounce inflation models, as well as the tensor-to-scalar ratio $r$, from the Planck+WP and BICEP2 data. For comparison, we also show the one dimensional constraints on $k_B$ and $r_B$ from the Planck+WP data alone (black dashed lines).\label{figure:2d}}
\end{center}
\end{figure}

Next, we consider the bounce inflation model. We use the Planck+WP data alone to constrain the parameters $k_B$ and $r_B$. We find the best fit values of $\log_{10}(k_B/{\rm Mpc}^{-1})={-2.6}$ and $r_B=0.8$ with the minimal $\chi^2(P)=9804.3$, which means the bounce model can only slightly improve the fit to the Planck data with $\Delta\chi^2\sim -1.4$. This result is slightly worse than some other works \cite{Liu:2013kea,Miranda:2013wxa,Miranda:2014wga}, due to our moderate suppression in the bounce model, which is shown in the left panel of figure \ref{figure:bestCMB}. In figure \ref{figure:2d}, we show the one-dimensional distributions on bounce parameters $k_b$ and $r_B$, and obtain the $95\%$ limits $\log_{10}(k_B/{\rm Mpc}^{-1}) < {-2.1}$ and $0 < r_B < 1$. The bounce model with no suppression is still consistent with the Planck+WP data. Again, similar to the $\Lambda$CDM, this bounce inflation model with $r=0$ can not fit the BICEP2 data as well, $\chi^2(B)=53.7$.
%due to the zero primordial CMB BB power spectrum.
%
Afterwards, we include the BICEP2 data and the tensor fluctuations into the analyses. Although in the bounce inflation, the theoretical CMB primordial BB power spectrum is suppressed at large scales, as shown in the right panel of figure \ref{figure:bestCMB}, the BICEP2 experiment can only measure the BB power spectrum at scales $\ell > 30$, where the suppression effect is very small. Therefore, the median value of the tensor to scalar ratio $r$ in the bounce inflation model is similar with that obtained in the standard $\Lambda$CDM model, $r=0.183\pm0.072$ at $95\%$ confidence level, as shown in figure \ref{figure:2d}. Meanwhile, we find that the suppression effect is obvious at very large scales $\ell < 20$. We expect that the Planck team will soon release the CMB polarization data which may cover the BB power spectrum at these scales. Therefore, it is very promising to examine the bounce inflation scenario in near future.

More importantly, adding the BICEP2 data significantly improves the constraints on parameters of the bounce inflation. The $68\%$ C.L. constraints are: $\log_{10}(k_B/{\rm Mpc}^{-1})={-2.4\pm0.2}$ and $r_B=0.71\pm 0.09$, while the $95\%$ limits are: $-2.8 < \log_{10}(k_B/{\rm Mpc}^{-1}) < -2.1$ and $0.54 < r_B < 0.88$. In figure \ref{figure:2d} we show the two-dimensional contours between $k_B$, $r_B$ and $r$. Since we have two free parameters to describe the suppression effect of the bounce model, when $k_b$ is increasing, the other parameter $r_B$ also becomes larger in order to compensate this effect. Therefore, the correlation of $r_B$ with $k_B$ is positive. On the other hand, the model with a non-zero $r$ brings the extra CMB TT power spectrum, which allows a large suppression, corresponding to an increasing $k_B$. So we find that there is a tiny positive correlation between $k_B$ and $r$.

Additionally, the $\chi^2$ values for the best fit model from the Planck+WP and BICEP2 data are $\chi^2(P)=9805.6$ and $\chi^2(B)=7.0$, respectively. The bounce inflation with $r=0.183$ can fit the BICEP2 data well, while it can also explain the Planck+WP data with the similar $\chi^2$ value, especially for the Planck low-$\ell$ TT data (see table \ref{chi2values}). The reason is that the extra CMB TT power spectrum at large scales, due to the non-zero tensor fluctuations, can be canceled by the suppression effect brought by the bounce, which is significantly different from the standard $\Lambda$CDM case. Based on these results, we conclude that when using Planck data alone, the bounce model can only slightly improve the fit to the data, comparing with the $\Lambda$CDM model. However, after including the BICEP2 data, the minimal $\chi^2$ becomes smaller in the bounce inflation model than that obtained in the standard $\Lambda$CDM case, $\Delta\chi_{min}^2\simeq -12$, corresponding to $\sim3.5\sigma$ confidence level. Based on the Akaike information criterion (AIC): ${\rm AIC} \equiv -2\ln{\mathcal{L}_{\rm max}}+2k$, where ${\mathcal{L}_{\rm max}}$ is the maximum likelihood achievable by the model and $k$ the number of parameters of the model \cite{aic}, we obtain the difference on the AIC between the standard inflation model and the bounce inflation model, $\Delta{\rm AIC}\equiv {\rm AIC}({\rm standard}) - {\rm AIC}({\rm bounce}) \simeq -8$. The bounce inflation model with two more parameters is strongly favored by the data and can very well fit to the Planck+WP and BICEP2 data simultaneously.

%Summary===============================================================
%\section{Summary}\label{Sum}

{\it Conclusions.---}Since a nonsingular bounce is expected to occur at an extremely high energy scale in very early universe, it is hard to detect directly by experiments. To search for a bounce, the associated observational consequences are significant in cosmological surveys. %This issue has been discussed widely in the literature, and one innovative clue is to study primordial perturbations. For example, in the context of the Pre-Big-Bang scenario \cite{Gasperini:1992em} and in the cyclic/Ekpyrotic model \cite{Khoury:2001wf}, the resulting cosmological perturbation was found to strongly depend on the physics at the epoch of thermalization \cite{Brustein:1994kn}, and the effective field approach to studying nonsingular bounces is proven to be very efficient and solid \cite{Cai:2012va}.
In the present Letter, we study the evolution of primordial gravitational waves in a combined scenario of matter bounce and inflation. We interestingly discover a novel jump feature on the power spectrum of these tensor modes at large scales, which could be verified by the Planck polarization data in the near future. The same feature was found to exist in the spectrum of primordial density perturbations. Importantly, this jump feature on the primordial scalar and tensor spectrum could alleviate problems of the excess power in the CMB temperature power spectrum.

Recently, the BICEP2 collaboration reports a $7\sigma$ detection of the non-zero tensor-to-scalar ratio, which corresponds to too large power in CMB TT power spectrum to fit in the standard $\Lambda$CDM framework. When we consider the bounce inflation model, the suppression effect could partially cancel those excess power at large scales. We perform a global analysis to constrain the jump features of both the scalar and tensor fluctuations from the Planck+WP and BICEP2 data. Our results reveal that the CMB data favor the bounce inflation model at about $3.5\sigma$ confidence level, namely $\log_{10}(k_{\rm B}/{\rm Mpc}^{-1})=-2.4\pm0.2$ (68\% C.L.) and $r_B = 0.71\pm0.09$ (68\% C.L.), when using Planck+WP and BICEP2 together. The bounce inflation model can simultaneously explain the Planck and BICEP2 data very well.

%Acknowledgments=======================================================
%\section*{Acknowledgements}

{\it Acknowledgments.---}We thank Si-Yu Li, Tao-Tao Qiu, Matteo Viel and You-Ping Wan for helpful discussions. J.X. is supported by the National Youth Thousand Talents Program. Y.C. is supported in part by NSERC and by the physics department at McGill University. H.L. is supported in part by the National Science Foundation of China under Grant Nos. 11033005 and 11322325, by the 973 program under Grant No. 2010CB83300. X.Z. is supported in part by the National Science Foundation of China under Grants Nos. 11121092, 11033005 and 11375202. The research is also supported by the Strategic Priority Research Program ``The Emergence of Cosmological Structures'' of the Chinese Academy of Sciences, Grant No. XDB09000000.

%End===================================================================

\end{document}